# Nano-Graphene Oxide for Cellular Imaging and Drug Delivery

Xiaoming Sun, Zhuang Liu, Kevin Welsher, Joshua Tucker Robinson, Andrew Goodwin, Sasa Zaric, Hongjie Dai*

*Department of Chemistry and Laboratory for Advanced Materials, Stanford University, Stanford, CA 94305, USA*

*\* Correspondence to hdai@stanford.edu*

Abstract

Two-dimensional graphene offers interesting electronic, thermal and mechanical properties that are currently explored for advanced electronics, membranes and composites. Here we synthesize and explore the biological application of nano-graphene oxide (NGO), i.e. single-layer graphene oxide sheets down to a few nanometers in lateral width. We develop functionalization chemistry to impart solubility and compatibility of NGO in biological environments. We obtain size separated pegylated NGO sheets that are soluble in buffers and serum without agglomeration. The NGO sheets are found to be photoluminescent in the visible and infrared regions. The intrinsic photoluminescence (PL) of NGO is used for live cell imaging in the near-infrared (NIR) with little background. We found that simple physisorption via π-stacking can be used for loading doxorubicin, a widely used cancer drug onto NGO functionalized with antibody for selective cancer cell killing *in vitro*. Owing to the small size, intrinsic optical properties, large specific surface area, low cost, and useful non-covalent interactions with aromatic drug molecules, NGO is a promising new material for biological and medical applications.

**Introduction.**

The interesting physical properties of graphene, a novel one-atom-thick two-dimensional graphitic carbon system, has led to much excitement in recent years in material science and condensed-matter physics[1-6]. Potential applications of graphene for nanoelectronics [1, 3], sensors and nanocomposites[4, 5] have been actively pursued[6]. The biological application of graphene and graphene oxide (GO) remains unexplored and wide-open. There are several prerequisites for biological applications for a new material. First, rational functionalization chemistry is needed to impart graphene aqueous solubility and biocompatibility. GO and its chemically converted derivatives form stable suspensions in pure water but aggregate in salt or other biological solutions[6]. Second, graphene sheets with suitable sizes are desired. Size control or size separation at various length scales is necessary to suitably interface with biological systems *in vitro* or *in vivo*. Graphene and GO samples thus far are typically microns or larger in size. Lastly, little is known experimentally about the properties of graphene with molecular dimensions, on the order of ~10nm or below. The optical properties of graphene and GO are largely unexplored, a topic of fundamental interest and could facilitate biological and medical research such as imaging.

Here, we prepared ultrasmall graphene oxide down to <10nm in lateral dimension. We imparted aqueous stability to the NGO in buffer solutions and other biological media by covalently grafting polyethylene glycol (PEG) star-polymers onto the chemically activated surfaces and edges. A rate separation method was developed to separate the pegylated NGO according to physical dimension or size. Interestingly, the NGO sheets showed photoluminescence from visible to the near-infrared (NIR) range, which was used for cellular imaging with little background. Furthermore, an aromatic anti cancer drug doxorubicin was loaded onto the NGO sheets at a high capacity via simple physisorption. The drug was selectively shuttled into cancer cells by antibody conjugated NGO for selective cancer cell killing.

**Results and Discussion**

Our preparation of ultrasmall graphene sheets started from graphene oxide (GO) made by using a modified Hummer's method (Fig. 1a, see Methods) [4-7]. Briefly, expandable graphite (Graftech Inc.) was used as starting material instead of graphite flakes for more uniform oxidization[3, 8]. The graphite powder was ground with NaCl salt crystals to reduce the particle size, and then soaked in sulfuric acid for 8h for intercalation. We then added oxidizing $KMnO_4$, increased the oxidization temperature and extended the oxidization time to ~2h to make fully and uniformly oxidized graphite (see Supp. Info.). The products were washed with dilute acid and water. The resulting GO after 1h sonication were mostly single layered (>70%, topographic height ~1.0nm) with 10-300nm in lateral width (Fig. 1b) according to AFM characterization. Previous work also attributed ~1.0nm thick GO to single-layered structures [3, 4, 6].

Infrared (IR) spectroscopy revealed the existence of –OH (~3400cm$^{-1}$), C=O (1715cm$^{-1}$), and C=C (1580cm$^{-1}$)[9, 10] functional groups on as made GO (Fig.1d). We activated the GO sample with chloroacetic acid in a strong basic condition to activate epoxide and ester groups, and to convert the hydroxyl groups to carboxylic acid (–COOH).[11] The intermediate product, named GO-COOH, had increased water solubility and more carboxylic acids for consequent PEGylation (see Methods). Upon grafting PEG stars (6-arm branched PEG molecules) onto the –COOH groups, we obtained a product (NGO-PEG) with high solubility and stability in salt and cellular solutions, which is desirable for biological applications. Without pegylation, GO and GO-COOH suspensions immediately aggregated in salt and any other biological solutions. Atomic force microscopy (AFM) observed mostly <20nm in sheet size of NGO-PEG (Fig.1c), while the as-made GO sheets were 10-300nm in size (Fig.1b). The ultra-small size of the NGO was caused by sonication involved in GO-COOH making and PEGylation steps. IR characterization on carefully purified NGO-PEG sample indicated strong -CH$_2$- (2870cm$^{-1}$) vibrations due to PEG chains, and characteristic amide-carbonyl (NH-CO) stretching vibration (~1650cm$^{-1}$, in Fig. 1d, labeled with an arrow.)[12], consistent with the grafting of PEG molecules on NGO sheets.

Activation and PEGylation of GO led to increases in optical absorption in the visible and near-infrared (VIS-NIR) range for the same starting graphitic carbon mass concentration (0.01mg/ml, Fig. 2a). The optical absorption peak at 230nm, originated from π-plasmon of carbon [13] [14], remained essentially the same. GO-COOH and NGO-PEG showed much higher absorbance in VIS-NIR range than GO. At 500nm, 808nm and 1200nm, the absorbance of GO-COOH and NGO-PEG were 480%, 780% and 470% of GO, respectively. The significant increase in absorbance led to solution color change (darkening), visible to the eye (Fig. 2a inset). Such darkening was observed in hydrazine reduction of graphene oxide, and attributed to restoring of the electronic conjugation within the graphene sheets [4, 6]. Here, we attributed the restoring to opening of epoxide groups and hydrolysis of esters on the GO in basic condition during the activation treatment. This led to local changes in the microstructures of NGO with released local strain and increased conjugation in the GO sheets, causing increased optical absorption in the Vis-NIR range.

The yellow brown color of our NGO solutions prompted us to investigate the photoluminescence of this material. Fluorescence measurement in the visible range revealed that the GO emission peaked at ~570nm at 400nm excitation (Fig. 2b). The emission maximum blue-shifted to ~520nm for NGO-PEG (Fig.2b). Chemical activation and PEGylation steps reduced the GO sheet size and changed the chemical functional groups on the sheets, which might be responsible for such shift. We probed into the IR and NIR regions and discovered photoluminescence (PL) of both GO and NGO-PEG in these regions (Fig. 2c&d). Fluorescence in the NIR is useful for cellular imaging due to little cellular auto-fluorescence in this region, as shown with single-walled carbon nanotubes[15, 16].

We developed a density gradient ultracentrifugation method [17, 18] to separate the NGO-PEG sheets by size (Supp. Info. Fig. S1) and gained insight to the photoluminescence properties of NGO. By making use of the different sedimentation rate of different sized graphene in a density gradient, and by terminating the sedimentation at suitable time points, we captured different sized graphene sheets at different positions along the centrifuge-tube (Fig.2e). AFM of different fractions clearly indicated size separation of NGO-PEG sheets by our method (Fig.2f-h.). However, to our surprise, the different sized NGO sheets exhibited similar optical absorbance, photoluminescence and PLE spectra (Supp. Info. Fig.S2), without apparent quantum confinement effects expected due to the different physical sizes of the separated NGO sheets.

This somewhat unexpected result suggested that small, conjugated aromatic domains existed on a NGO sheet. That is, small conjugated domains with various sizes (~1-5nm) coexist in a single, physically connected NGO sheet. Indeed, careful AFM imaging found small domain-like structures 1-5nm in size (Supp. Info. Fig.S3). Separation of NGO sheets by physical size afforded various fractions exhibiting similar photoluminescence since the NGO-PEG sheets contained similar smaller aromatic domains. The domain sizes were inhomogeneous and ranged from small aromatic molecules to large macromolecular domains. The former was responsible for fluorescence in the visible range, while the latter gave PL in IR range. The existence of conjugated aromatic domains spaced by non-conjugated aliphatic six-members ring structures were found on GO by previous NMR experiments [19]. Nevertheless, our proposed photoluminescence mechanism is merely a suggestion, and requires further investigation. The resolution of size separation may also need significant improvement to observe quantum confinement effect, especially at the low end of the size distribution. Note that previously, carbon nanoparticles (1-5nm) after oxidative acid treatment exhibited fluorescence in the visible, and particle size and surface states significantly affected the fluorescence intensity and peak positions [20, 21]. The mechanism and chemical species responsible for the fluorescence also remained unclear [20, 21].

Fluorescent species in the NIR and IR range are potentially useful for biological applications since cells and tissues exhibit little auto-fluorescence in this region[16]. To this end, we covalently conjugated a B-cell specific antibody Rituxan (anti-CD20) to NGO-PEG (NGO-PEG-Rituxan) to selectively recognize and bind to B cell lymphoma cells (Fig. 3a)[16]. We incubated B-cells and T-cells in solutions of NGO-PEG-Rituxan conjugates at 4 °C to allow the conjugates to interact with the cell surface but block internalization via endocytosis [22]. The cells were then washed and imaged by detecting NIR photoluminescence from 1100 to 2200 nm using an InGaAs detector under 658 nm laser excitation (laser spot size ~1μm, see Methods). We detected the intrinsic NIR photoluminescence of NGO-PEG selectively on positive Raji B-cells surface (Fig.3b) and not on negative CEM T-cells (Fig.3c). This confirmed selective NGO-PEG-Ab binding to B-cells over T cells (Fig.3d). It also established NGO as a NIR fluorophore for selective biological detections and imaging to take advantage of the little cellular auto-fluorescence in the NIR region [16, 23].

The fluorescence quantum yield (QY) of the NGO-PEG was difficult to quantify due to the inhomogeneous species in the sample. As a preliminary attempt, we compared the total emitted light by Hipco SWNTs [24-26] and NGO-PEG in the 900-1500 nm region under 785nm excitation, and observed similar light emission (within an order of magnitude) (Supp. Info. Fig. S4) normalized to the same absorbance at 785nm. Notably that for SWNTs, the quantum yield issue is also not well resolved due to sample inhomogenity, and is believed to be up to several percent for certain chiralities[24, 26]. Establishing the quantum yield of graphitic nanomaterials requires significant future effort including the improvement of material homogeneity and/or separation.

Next, we explored using NGO as sheet-like vehicles to transport an aromatic anticancer drug doxorubicin (DOX) into cancer cells[27]. We first checked if NGO-PEG exhibit any toxicity by incubating Raji cells in various concentrations of NGO-PEG for 72 h. We observed slight reduction of cell viability only for extremely high NGO-PEG concentrations (>100mg/L) (supp. Info. Fig. S5). In our subsequent cellular experiments, we used only ~2mg/L of NGO-PEG, a concentration far below the level to cause any appreciable toxic effect to the cells.

Rituxan (CD20+ antibody) conjugated NGO-PEG was used to target specific cancer cells for selective cellular killing (Fig.4a). The loading of DOX, a widely used chemotherapy drug for treating various cancers, was performed by simple mixing of a NGO-PEG-Ab solution with DOX at pH 8 overnight, followed by repeated filtering to remove free, unbound DOX in solution. The formation of NGO-PEG/DOX was visible from the reddish color of the NGO-PEG/DOX solutions due to adsorbed DOX and its characteristic UV–vis absorbance peak at 490 nm on top of the NGO-PEG absorption spectrum (Fig.4b and inset). The loading of DOX onto NGO was attributed to simple π-stacking, similar to that in carbon nanotubes[27]. Under AFM, obvious thickness increase was observed as DOX was stacked onto graphene sheets (Supp. Info. Fig.S6). Compared to single-walled carbon nanotubes for drug loading via π-stacking[27], NGO is advantageous in low cost and high scalability[27].

Drug release from NGO-PEG sheets was observed as the chemical environment changed to acidic conditions. We found that ~40% of DOX loaded on NGO-PEG was released over 1 day in an acidic solution of pH 5.5 (Fig.4c), which was attributed to the increased hydrophilicity and solubility of DOX at this pH[27]. The release rate was slowed down as the pH was adjusted to pH 7.4, ~15% over 2 days. The pH-dependent drug release from NGO-PEG could be exploited for drug delivery applications since the micro-environments in the extracellular tissues of tumors and intracellular lysosomes and endosomes are acidic, which will afford active drug release from NGO-PEG delivery vehicles.

For DOX loaded onto NGO-PEG-Rituxan, we incubated the conjugates with Raji cells at 2μM and 10μM DOX concentrations. Enhanced DOX delivery and cell killing was evidenced by comparison with Raji cells treated by free DOX, NGO-PEG/DOX without Rituxan, and a mixture of NGO-PEG/DOX and Rituxan without covalent binding (Fig.4d). It should note that the *in vitro* selectivity is limited by the passive uptake of NGO-PEG/DOX. At higher DOX concentrations, the passive uptake is high so that cells are killed by either NGO-PEG/DOX or NGO-PEG-Rituxan/DOX. However a minimal DOX concentration is required to exhibit any biological effect. Therefore there is a dosage window in which the selective cell killing can be achieved. This is common for many other targeting drug delivery systems. We observed very good selective at [DOX]=10μM. The percentage of cell growth inhibition increased from ~20% in the NGO-PEG/DOX case to ~80% in the NGO-PEG-Rituxan/DOX case, which was a significant enhancement [27]. This result demonstrated the potential of selective killing of cancer cells using NGO-PEG-antibody/drug conjugates *in vitro*.

**Conclusions**

In summary, multifunctional biocompatible nano-graphene oxides with various physical sizes were prepared in a scalable manner. Photoluminescence of NGO from visible through infrared range was revealed and used for cellular imaging. Anti cancer drug was loaded onto NGO with high capacity, and selectively transported into specific cancer cells by antibody guided targeting. The novel graphitic nanostructures, combined with multi-functionalities including biocompatibility, photoluminescence and drug loading and delivery, suggest promising applications of graphene materials in biological and medical areas.

**Acknowledgements**

This work was supported by NIH-NCI funded CCNE-TR at Stanford. We are grateful to Drs. Alice Fan, and Dean Felsher for providing the antibodies used in this work.

**Electronic supplementary materials**

Experimental details on synthesis, Pegylation of NGO, characterizations, antibody (Rituxan) conjugation, cell culture, and NIR imaging of cells can be found in the supplementary materials with 6 supplementary figures.

## METHODS

### SYNTHESIS OF GRAPHENE OXIDE (GO)

Graphene oxide (GO) was made by modified Hummer's method[7] using expandable graphite flake (Graftech) as starting material[3]. 1g Expandable graphite flake was ground with 50g NaCl solids for 10 min. NaCl was dissolved and removed by filtration with water (lose ~15% carbon during this step). Ground expandable graphite flake (~0.85 g) was stirred in 23ml $H_2SO_4$ (98%) for 8h. 3g $KMnO_4$ was gradually added while keeping temperature <20ºC. The mixture was stirred at 35-40ºC for 30 min, then at 65-80ºC for 45min. Next, 46ml water was added and heated at 98-105ºC for 30 min. The reaction was terminated by addition of 140ml distilled water and 10ml 30% $H_2O_2$ solution. The mixture was washed by repeated centrifuge and filtration, first with 5% HCl aqueous solution, then DI water. 160ml water was added to the final product and vortexed well to make an even suspension for storage.

### PEGYLATION OF GRAPHENE OXIDE

For pegylation, 5ml GO aqueous solution was diluted by 2, and then bath sonicated for 1h to make a clear solution. 1.2g NaOH and 1.0g chloricacetic acid (Cl-$CH_2$-COOH) was added into the 10ml GO suspension (~2mg/ml) and bath sonicated for 1-3 h [11] to convert the –OH groups to –COOH via conjugation of acetic acid moieties (named GO-COOH). The obtained GO-COOH solution was neutralized, and purified by rinsing and filtrations. GO-COOH suspension was diluted by water to make into optical density OD=0.4 at 808nm (1mm optical path). 2mg/ml 6-arm Polyetheylene Glycol-Amine (Sunbio Inc.) was added into GO-COOH suspension and sonicated for 5min. N-(3-Dimethylaminopropyl-N'-ethylcarbodiimide hydrochloride (EDC, from Sigma Inc.) was added twice to reach 4mM and reacted overnight, then quenched by Mercaptoethanol (Fluka Inc.). The final product (NGO-PEG) was obtained by ultracentrifugation at 45k rpm in 2× phosphate buffered saline (PBS) for 1h to save the supernatant (yield ~50%). The aggregates were discarded.

**SEPARATION**

Rate separation in step density was used for NGO-PEG separation. In a typical experiment, OptiPrep® (60% (w/v) iodixanol, 1.32 g·cm$^{-3}$) (Sigma-Aldrich Inc.) was diluted with water to make 5%, 10%, 15%, and 20% iodixanol solutions. Gradients were created directly in Beckman centrifuge tubes (polycarbonate, inner diameter 13 mm, length 51 mm) by adding the 4 layers (0.6 mL each) to the tube bottom sequentially in the order of increasing density. Finally, 0.4 mL of 60% iodixanol was added to the bottom of the centrifuge tube to raise the height of the gradient in the centrifuge tube. 0.2 mL freshly made NGO-6PEG was layered on top of density gradient prior to ultracentrifugation. The typical centrifugation condition was 2.5 h at 50k round per minute (~300kg). Calibrated micropipettors were used to manually sample 100 μL fractions along the centrifuge tube after ultracentrifugation.

**CHARACTERIZATION**

AFM was used for graphene oxide sheet size and thickness characterization. The optical properties of graphene oxide was characterized by ultraviolet-visible-near-infrared absorbance spectrometer (Varian Cary 6000i), fluorescence spectrometer (Spex Fluorolog 3, $\lambda_{ex}$ = 400nmnm) and home built photoluminescence-excitation (PLE) spectrometer at room temperature [16].

**CELLULAR IMAGING**

Thiolated Rituxan was conjugated to the amine groups on NGO-PEG via a sulfo-SMCC linker (Pierce Inc.)[16]. For the cell incubation, 200 μl of CEM.NK T-cell and Raji B-cell (~1 million/ml) were incubated with 50 μl of NGO-PEG with or without Rituxan conjugation in PBS for 1h at 4°C. The NGO-PEG concentration in the solution during incubation was ~0.7mg/ml. Cells were washed with PBS 3 times to remove unbound NGO-PEG before use for NIR photoluminescence imaging. Cell samples prepared as described above were placed in a sample holder with a thin 200μm quartz window. All NIR fluorescence images were taken using an inverted NIR fluorescence microscope in confocal mode. Excitation from a diode laser at 658 nm (Renishaw) was focused using a 100× IR coated objected lens (Olympus). The laser spot size width on the sample was about 1μm FWHM. The laser intensity at the sample was ~20mW. Emitted light was collected through the same objective and focused onto an OMA-V 1024 element linear InGaAs array (Princeton Instruments). The collected light went through a 900nm long pass filter (Omega) and a 1100nm long pass filter (ThorLabs) to block reflected excitation and reduce background fluorescence from the sample holder. High resolution images were taken by inserting a 50μm pinhole in the collection path, and 1 micron steps were taken in 2 directions. Background fluorescence from the sample holder (~160 counts) was subtracted to give relevant statistics about the effectiveness of NGO-PEG binding.

**DRUG LOADING AND CELLULAR TOXICITY**

Doxorubicin (DOX) loading onto NGO-PEG (and NGO-PEG-Rituxan) was done by simply mixing 0.5mM of DOX with the NGO-PEG solution (~0.2mg/ml) at pH 8 overnight. Unbound excess DOX was removed by filtration through a 100kDa filter and repeated rinse. The formed NGO-PEG/DOX complexes were re-suspended and stored at 4°C. Concentration of DOX loaded onto NGO-PEG was measured by the absorbance peak at 490 nm (characteristic of DOX, after subtracting the absorbance of NGO-PEG at that wavelength) with a molar extinction coefficient of $1.05 \times 10^4$ M·cm$^{-1}$. Both Raji and CEM cells were incubated with free DOX, NGO-PEG/DOX, NGO-PEG/DOX + Rituxan (unconjugated) and NGO-PEG-Ri/DOX at DOX concentrations of 2 μM or 10 μM for 2 hours and washed by PBS twice before transferring into fresh cell medium. After another 48h incubation, Cell viability was measured by the MTS assay with CellTiter96 kit (Promega).

**FIGURE CAPTION**

**Figure 1.** Synthesis and pegylation of Nano- Graphene Oxide. (a) Schematic illustration of pegylation of graphene oxide by PEG-stars; (b)&(c) AFM images of GO and NGO-PEG, respectively; (d) IR spectra of GO, GO-COOH and NGO-PEG. (the black arrow indicated characteristic amide-carbonyl vibration.).

**Figure 2.** Optical properties of nano- graphene oxide sheets. (a) UV-VIS-NIR absorbance spectra of GO, GO-COOH and NGO-PEG solutions with 0.01mg/ml graphitic carbon (1cm optical path). Inset: a photograph of GO and NGO-PEG solutions at the same graphitic carbon concentration to show visible color difference. (b) Fluorescence of GO (black curve) and NGO-PEG (red curve) in the visible range under an excitation of 400nm. (c)&(d) Photoluminescence excitation (PLE) spectra of GO & NGO-PEG with 0.31mg/ml graphitic carbon in the IR region (1mm optical path). The emission intensity at various emission wavelengths (x-axis) is represented by the color scheme shown as a function of excitation wavelength (y-axis). The data were obtained after normalization to detector sensitivity and absorbance curves. (e) A photograph of a ultracentrifuge-tube after density gradient separation of NGO-PEG. (f)-(h) AFM images of NGO-PEG fractions (F5, F13 and F23) at different locations in the centrifuge-tube as labeled in (e).

**Figure 3.** Nano-graphene for targeted NIR imaging of live cells.
(a) A schematic drawing illustrating the selective binding and cellular imaging of

NGO-PEG conjugated with anti-CD20 antibody, Rituxan; (b) NIR fluorescence image of CD20 positive Raji B-cells treated with the NGO-PEG-Rituxan.conjugate. Scale bar shows intensity of total NIR emission (in the range of 1100-2200 nm). Images are false-colored green. (c) NIR fluorescence image of CD20 negative CEM T-Cells treated with NGO-PEG-Rituxan conjugate. (d) Mean NIR fluorescence intensities in the image area for the both the positive (Raji) and negative (CEM) cells treated by NGO-PEG-Rituxan conjugate.

**Figure 4.** Nano-graphene oxide for target drug delivery. (a) A schematic illustration of Doxorubicin (DOX) loading onto NGO-PEG-Rituxan via π-stacking. (b) UV-VIS-NIR absorbance spectra of NGO-PEG and NGO-PEG/DOX. DOX loading on NGO-PEG was evidenced by a strong absorption peak centered at ~490nm. Reddish color of the DOX loaded NGO-PEG is seen in the solution (see inset); (c) Doxorubicin retained on NGO-PEG over time in buffers at pH 5.5 and 7.4; (d) *In vitro* toxicity test at 2μM and 10μM DOX concentration to show Rituxan selectively enhanced doxorubicin delivery into Raji B-cells by comparing NGO-PEG-Rituxan/DOX with free DOX, mixture of DOX with NGO-PEG, and mixture of DOX, Rituxan and NGO-PEG. The viable cell percentage was measured by the MTS assay.

**Figure 1**

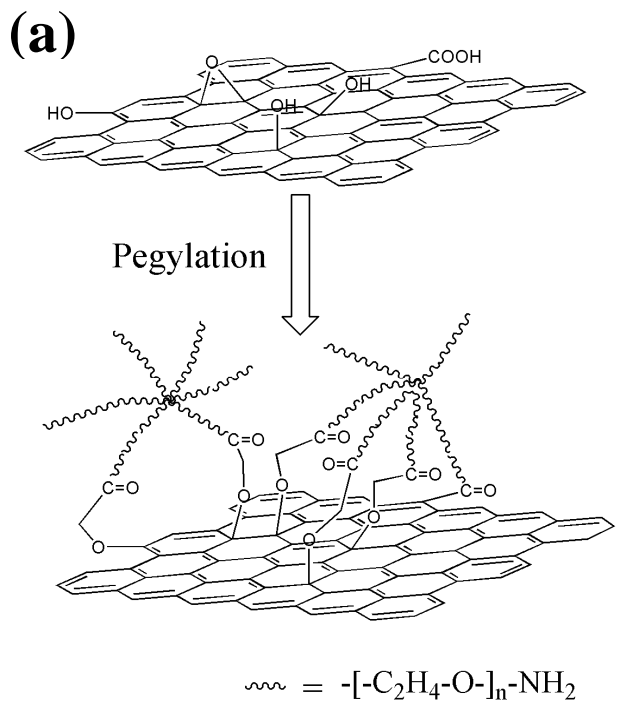

$\sim\sim\sim$ = -[-C$_2$H$_4$-O-]$_n$-NH$_2$

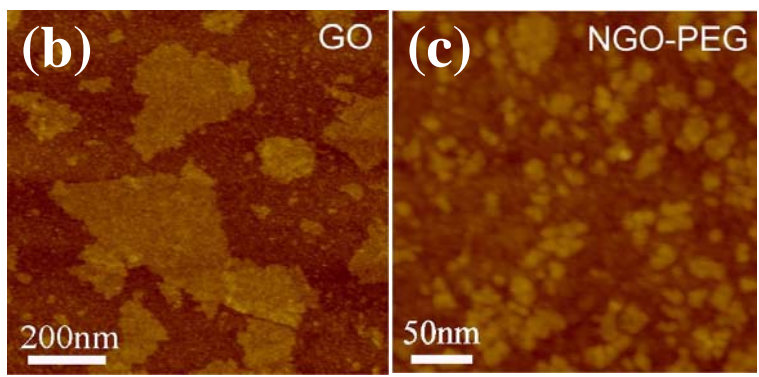

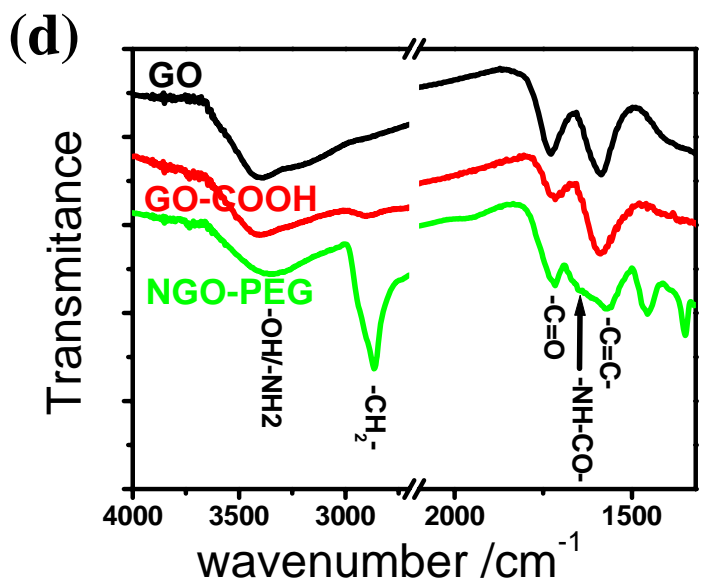

Figure 2

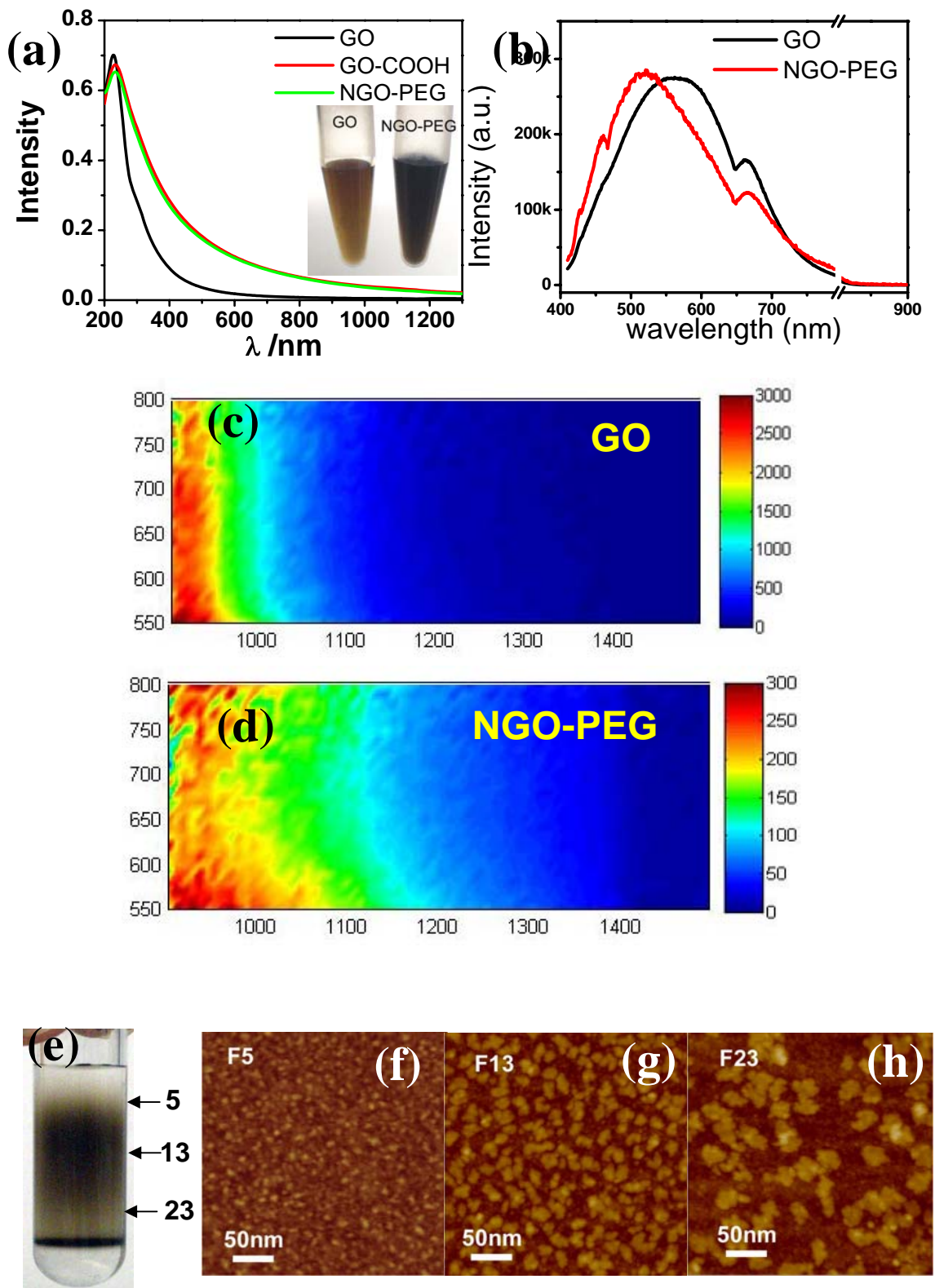

Figure 3

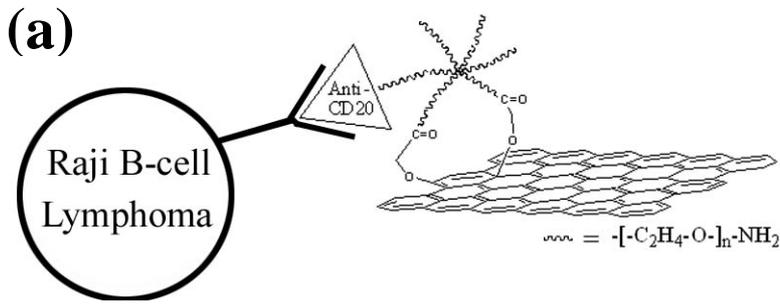

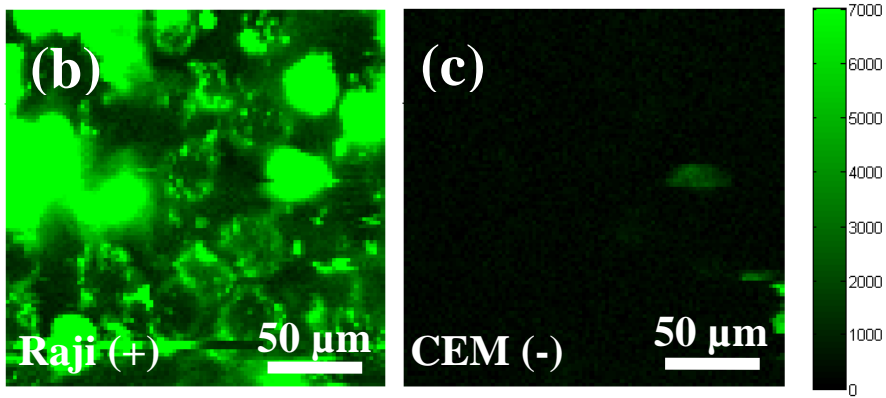

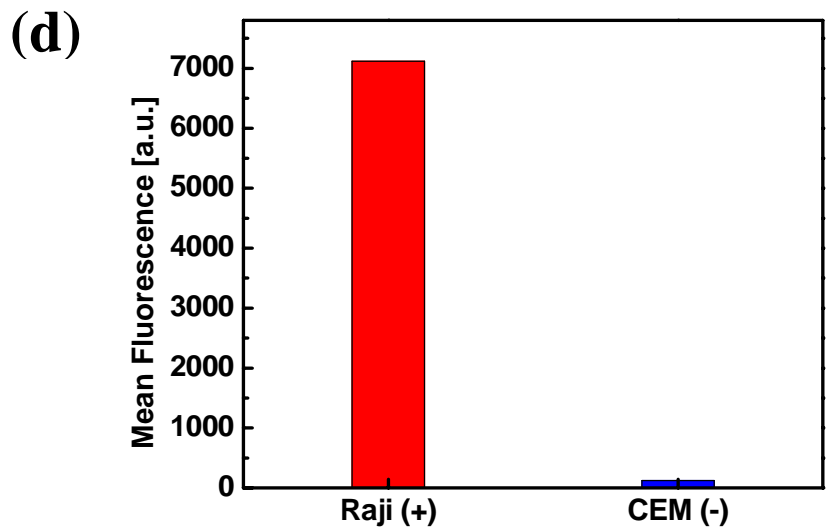

Figure 4

# Nano-Graphene Oxide for Cellular Imaging and Drug Delivery

Xiaoming Sun, Zhuang Liu, Kevin Welsher, Joshua Tucker Robinson, Andrew Goodwin, Sasa Zaric, Hongjie Dai\*

*Department of Chemistry and Laboratory for Advanced Materials, Stanford University, Stanford, CA 94305, USA*

*\* Correspondence to hdai@stanford.edu*

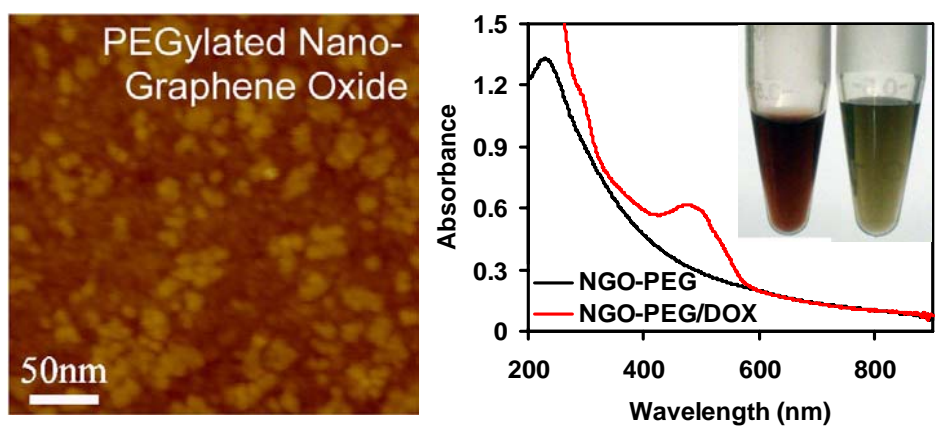

**PEGylated graphene oxide with lateral dimension <10nm was obtained to afford solubility and stability in biological solutions including serum, and used for near-infrared cellular imaging and targeting drug delivery *in vitro*.**

\*\*\*\*\*\*\*\*\*\*\*\*\*\*\*\*\*\*\*\*\*\*\*\*\*\*\*\*\*\*\*\*\*\*\*\*\*\*\*\*\*\*\*\*\*\*\*\*\*\*\*\*\*\*\*\*\*\*\*\*\*\*\*\*\*\*